# Convolutional Neural Network-Aided Bit-Flipping for Belief Propagation Decoding of Polar Codes

*Chieh-Fang Teng, Andrew Kuan-Shiuan Ho, Chen-Hsi (Derek) Wu, Sin-Sheng Wong, and An-Yeu (Andy) Wu*

*Abstract*—Known for their capacity-achieving abilities, polar codes have been selected as the control channel coding scheme for 5G communications. To satisfy the needs of high throughput and low latency, belief propagation (BP) is chosen as the decoding algorithm. However, in general, the error performance of BP is worse than that of enhanced successive cancellation (SC). Recently, critical-set bit-flipping (CS-BF) is applied to BP decoding to lower the error rate. However, its trial and error process result in even longer latency. In this work, we propose a convolutional neural network-assisted bit-flipping (CNN-BF) mechanism to further enhance BP decoding of polar codes. With carefully designed input data and model architecture, our proposed CNN-BF can achieve much higher prediction accuracy and better error correction capability than CS-BF but with only half latency. It also achieves a lower block error rate (BLER) than SC list (SCL).

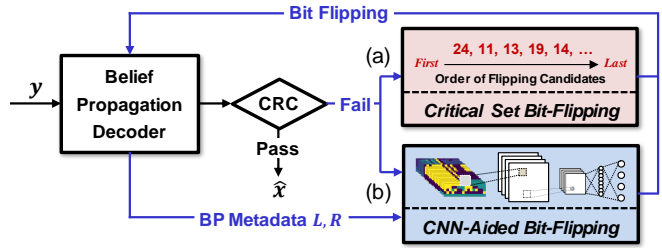

Fig. 1. The mechanisms of bit-flipping for polar decoder: (a) critical set bit-flipping (CS-BF) [15], and (b) proposed convolutional neural network-aided bit-flipping (CNN-BF).

## I. INTRODUCTION

Polar code is a type of block channel code proven to achieve channel capacity first proposed by Arikan [1]. In recent years, it has received intensive attention due to its adoption as the enhanced mobile broadband (eMBB) control channel coding scheme for 5G New Radio (NR) by 3GPP [2].

The main two algorithms for polar decoding are successive cancellation (SC) and belief propagation (BP). Compared with BP decoding, SC decoding can fulfill the channel-capacity ability and achieve a lower block error rate (BLER) through enhanced SC algorithm [3]-[4]. However, SC suffers from low throughput due to its sequential processing nature, while BP algorithm excels in architectural parallelization, thus has lower decoding latency [5]. Recently, considerable efforts have been put into improving the error performance of BP to achieve that of SC, while still maintaining its advantages. One method for BP performance optimization is the inclusion of neural networks to assist with BP decoding process [6]-[9]. In [6]-[9], the BP algorithm is enhanced through the scaling of messages from trainable weights. It reduces the total number of iterations before convergence and overall complexity, but does not address the lacking error correction performance.

In order to lower the BLER of decoders, in general, bit-flipping (BF) decoders can be utilized to minimize error propagation, by performing error corrections on incorrectly decoded bits during decoding iterations [10]-[15]. In [10]-[14], although BF has been successively applied to SC, it still inherently suffers from longer decoding latency. On the other hand, with the aid of BF, BLER of BP-based decoding can be comparable with that of SC-based algorithms [15]. However, two issues should be addressed:

1) *Increase in decoding latency*: The principle of BF is to generate many probable candidates, and perform iterations until a condition is satisfied as shown in Fig. 1(a). Therefore, the latency is directly correlated to the number of cases that fail. This trial-and-error method is especially problematic in worst-case scenarios, where the latency is significantly greater than the non-BF approach, by a factor of the number of tries.
2) *Sub-optimal search space*: As described in [15], the critical set (CS), a subset of high-risk information bits, is adopted to perform BF operation. Although it can effectively limit the search space, it also leads to some uncorrectable errors, thus degrading the error correction capability of BF.

In this paper, by taking advantage of the emerging deep learning (DL) techniques, we propose a novel convolutional neural network-aided bit-flipping (CNN-BF) polar decoder as shown in Fig. 1(b), which provides a flexible adjustment between decoding performance and latency. Our main contributions are summarized as below:
1) The process of bit-flipping candidate selection is replaced by the CNN model to exploit the variation of BP decoding process and dynamically identifies the erroneous bit. It can achieve the same BLER but with only half flipping attempts compared with the state-of-the-art critical-set bit-flipping (CS-BP) algorithm [15].
2) The input data for the CNN model is carefully designed, which is transformed from the metadata of the BP decoding process. Furthermore, domain-specific data pre-processing is adopted to reduce model complexity and increase prediction accuracy.

The rest of this paper is organized as follows. Section II briefly reviews BP and CS-BF decoders. Section III illustrates the input data with the proposed CNN architecture and its integration into the BF process. The numerical experiments and analyses are shown in Section IV. Finally, Section V concludes our work.

C.-F. Teng, S.-S. Wong, and A.-Y. A. Wu are with the Graduate Institute of Electronics Engineering, National Taiwan University, Taipei, Taiwan (e-mail: {jeff, shawn, andywu}@access.ee.ntu.edu.tw).

A. K.-S. Ho and C.-H. D. Wu are with the Department of Electrical Engineering, National Taiwan University, Taipei, Taiwan (e-mail: {b04901167, b03901054}@ntu.edu.tw).

This research work is financially supported by the MediaTek Inc., Hsinchu, Taiwan, under Grants MTKC-2019-0070. The first author is also sponsored by MediaTek Ph.D. Fellowship program.

## II. POLAR CODES AND PRIOR WORKS

### A. Polar Codes with Belief Propagation Decoding

To construct an $(N, K)$ polar codes, the $N$-bit message $\boldsymbol{u}^N$ is recursively constructed from a $2 \times 2$ polarizing transformation $\boldsymbol{F} = \begin{bmatrix} 1 & 0 \\ 1 & 1 \end{bmatrix}$ by $\log_2 N$ times to exploit the channel polarization [1]. As $N \to \infty$, the synthesized channels tend to two extremes: the noisy channels (unreliable) and noiseless channels (reliable). Therefore, the $K$ information bits are first assigned to the $K$ most reliable bits in $\boldsymbol{u}^N$ and the remaining $(N - K)$ bits are referred to as frozen bits with the assignment of zeros. Then, the $N$-bit transmitted codeword $\boldsymbol{x}^N$ can be generated by multiplying $\boldsymbol{u}^N$ with generator matrix $\boldsymbol{G}_N$ as follows:

$$\boldsymbol{x}^N = \boldsymbol{u}^N \boldsymbol{G}_N = \boldsymbol{u}^N \boldsymbol{F}^{\otimes n} \boldsymbol{B}_N, n = \log_2 N. \quad (1)$$

$\boldsymbol{F}^{\otimes n}$ is the $n$-th Kronecker power of $\boldsymbol{F}$ and $\boldsymbol{B}_N$ represents the bit-reversal permutation matrix.

Belief propagation (BP) is a widely used message passing algorithm for decoding, such as low-density parity-check (LDPC) codes and polar codes. The decoding process of polar codes is to iteratively apply BP algorithm over the corresponding factor graph as shown in Fig. 2. For an $(N, K)$ polar codes, there are $n = \log_2 N$ stages and total $N \times (n + 1)$ nodes on the factor graph. Each node $(i, j)$ represents $j$-th node at the $i$-th stage in the factor graph. It has two types of log likelihood ratios (LLRs), namely left-to-right message $R_{i,j}^{(t)}$ and right-to-left message $L_{i,j}^{(t)}$, where $t$ represents the $t$-th iteration. Before beginning iterative propagation and the updating of node values, their LLR values are first initialized as:

$$\begin{cases} R_{0,j}^{(1)} = \begin{cases} 0, & if\ j \in A \\ +\infty, & if\ j \in A^c \end{cases} \\ L_{n,j}^{(1)} = \ln \dfrac{P(y_j|x_j = 0)}{P(y_j|x_j = 1)} \end{cases}, \quad (2)$$

where $A$ and $A^c$ are the set of information bits and the set of frozen bits, respectively.

Then, the iterative decoding procedure with the updating of $R_{i,j}^{(t)}$ and $L_{i,j}^{(t)}$ is given by:

$$\begin{cases} L_{i,j}^{(t)} = g\left(L_{i+1,j}^{(t)}, L_{i+1,j+N/2^i}^{(t)} + R_{i,j+N/2^i}^{(t)}\right), \\ L_{i,j+N/2^i}^{(t)} = g\left(R_{i,j}^{(t)}, L_{i+1,j}^{(t)}\right) + L_{i+1,j+N/2^i}^{(t)}, \\ R_{i+1,j}^{(t)} = g\left(R_{i,j}^{(t-1)}, L_{i+1,j+N/2^i}^{(t)} + R_{i,j+N/2^i}^{(t)}\right), \\ R_{i+1,j+N/2^i}^{(t)} = g\left(R_{i,j}^{(t)}, L_{i+1,j}^{(t-1)}\right) + R_{i,j+N/2^i}^{(t)}, \end{cases} \quad (3)$$

where $g(x, y) \approx \text{sign}(x)\text{sign}(y)\min(|x|, |y|)$ is the min-sum approximation introduced to reduce complexity. Finally, after $T$ iterations, the estimation of $\hat{\boldsymbol{u}}^N$ is decided by:

$$\hat{u}_j^N = \begin{cases} 0, & if\ L_{0,j}^{(T)} + R_{0,j}^{(T)} \geq 0, \\ 1, & if\ L_{0,j}^{(T)} + R_{0,j}^{(T)} < 0. \end{cases} \quad (4)$$

### B. Prior Work: Critical Set Bit-Flipping (CS-BF) Belief Propagation Decoder [15]

Bit-flipping (BF) is an assistive mechanism to the decoding process, where a possibly incorrectly decoded bit is guessed and flipped prior to the restarted decoding process. Thus, precise BF can effectively improve the block error rate (BLER) performance for polar codes. Due to the message passing algorithm of BP decoding, the incorrect decoding of information bits may result in error propagation and thus negatively affect the reliability and accuracy of many other bits. To address the issue of error propagation, the mechanism of BF flips the value of previous estimated $\hat{u}_j^N$, and sets the a priori knowledge of $u_j^N$ to infinity as if it is a frozen bit. Therefore, the initialized values of $\boldsymbol{R}_0$ in (2) are revised as:

$$R_{0,j}^{(1)} = \begin{cases} 0, & if\ j \in \{A \backslash F\} \\ \infty \times (2\hat{u}_j^N - 1), & if\ j \in F \\ +\infty, & if\ j \in A^c \end{cases}, \quad (5)$$

where $F$ is the set of flipping positions. By doing so, the a priori knowledge of flipped bits is expected to correct the other wrongly propagated messages in the previously failed BP decoding.

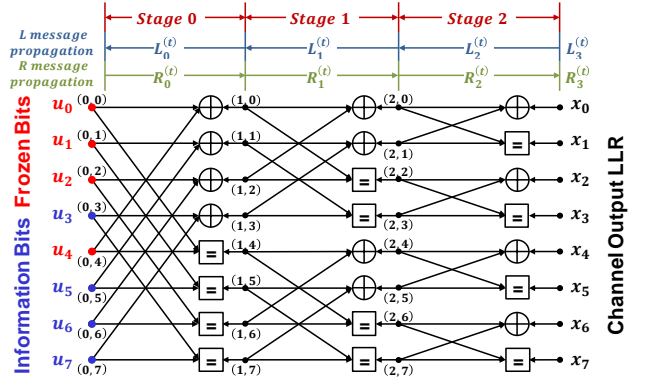

Fig. 2. Factor graph of polar codes with $N = 8$. $A = \{3, 5, 6, 7\}$ and $A^c = \{0, 1, 2, 4\}$.

According to the algorithm detailed above, the decoding latency of BF corresponds to the required number of flipping attempts, which is dominated by the correction of flipped bits. Therefore, critical set (CS), consisting most of the error-prone bits, was proposed in [10]-[15]. Now only the bits in CS are considered for flipping. The critical set is constructed based on the structure of polar code, where the first nodes in subtrees of all information bits are high risk, and thus are included in the set. By only selecting bits for BF from CS, it results in less flipping attempts and achieves lower latency.

With the established CS, the conventional BP decoding process can commence. If BP fails to decode successively, checked by a cyclic redundancy check (CRC), a candidate bit from the critical set is selected for flipping according to Eq. (5) as shown in Fig. 1(a). After bit-flipping, BP decoding is performed again. If the result satisfies CRC, the decoding process is completed. Otherwise, the other candidates in the critical set are attempted until CRC is successfully passed. In this work, we mainly focus on 1-bit correctable codewords, which can be successfully decoded with only one correct BF. This is the same case for $\omega = 1$ as in [15]. For more details about bit-flipping and critical set, please refer to [10]-[15].

## III. PROPOSED CONVOLUTIONAL NEURAL NETWORK-AIDED BIT-FLIPPING DECODER

### A. BP Metadata with Pre-Processing of Input Data

Although the proposed CS-BF in [15] can reduce the number of flipping attempts and the decoding latency, two issues need to be addressed. Firstly, this mechanism is essentially still a process of trial-and-error to attempt all the bits in critical set. As the search space in critical set grows up with block length $N$, it still results in intolerable decoding latency. Secondly, though the critical set can effectively limit the search space from all information bits to a smaller subset, it does not include error bits outside the critical set, thus degrading the error correction capability of BF. Therefore, we propose a model-based approach to predict the flipping position, which results in less flipping attempts and achieves better error correction capacity.

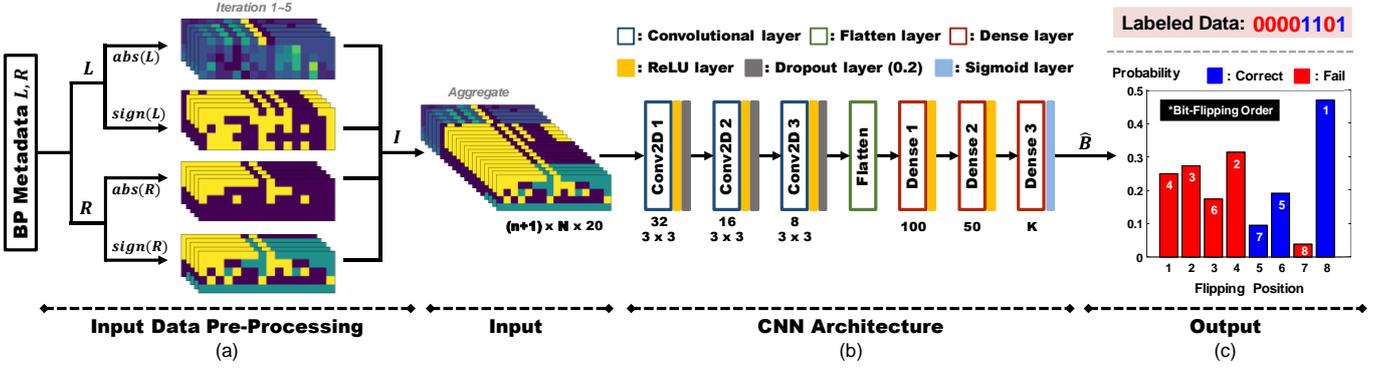

Fig. 3. The detailed overview of the proposed convolutional neural network-aided bit-flipping decoder: (a) illustration of the process of input data pre-processing; (b) proposed convolutional neural network architecture; (c) illustration of labeled data and bit-flipping order based on model's prediction results.

The input data for the model-based approach is important since it has a significant impact on prediction accuracy. In our case, we make use of the metadata from BP, namely the values of LLRs $L, R$ on the factor graph, as the input data for the training and prediction. In each BP decoding iteration, we record the values of LLRs on the whole factor graph and map the values to an image as shown in Fig. 3(a). Also, due to the iterative decoding process, the images, representing different iterations, will be jointly integrated as input data. Therefore, the adopted model can explore not only the relation between connected nodes but also the variation of LLRs among different iterations, namely in both spatial and temporal dimensions.

In addition, to further improve the prediction accuracy and reduce model complexity, we apply some domain-specific signal pre-processing before feeding the input data into the model. Two features, the absolute and sign values, are extracted from LLRs as shown in Fig. 3(a). By doing so, the absolute values represent the reliability of each node and the sign values are helpful for the model to further explore the variation between different nodes.

### B. Architecture Design of Convolutional Neural Network

Suppose that the number of iterations for BP is 5, there are total 20 images after data pre-processing with each image resolution being $(n+1) \times N$, which is consistent with the size of the factor graph. For the image-based input data $I$, the convolutional neural network (CNN) is employed as it is the widely used model for image processing with the ability to extract local connectivity and subtle features of the input image. The architecture of the proposed CNN model is illustrated in Fig. 3(b). It is constructed by three two-dimensional convolutional layers, followed by three dense layers. The values below the convolutional layer represent the number and size of filters, respectively. On the other hand, the values below the dense layer represent the number of nodes. Moreover, the nonlinear activation function, Rectified Linear Units (ReLUs), among each layer is defined as:

$$f_{ReLU}(x) = \max\{0, x\}. \qquad (6)$$

It is helpful for extracting more complex features. Besides, to reduce overfitting, the regularization technique of "dropout"

**Algorithm 1:** Proposed Convolutional Neural Network-Aided Bit-Flipping Decoder

**Input:** $y$, $A$, $T_{max}$
**Output:** $\hat{u}^N$
1: $L, R \leftarrow$ initialize the BP decoder using (2)
2: $\hat{u}^N, L, R \leftarrow$ BP decoder($L, R$)
3: $t \leftarrow 1$
4: **if** $\hat{u}^N$ does not pass CRC **do**
4:     $I \leftarrow$ input data pre-processing($L, R$)
5:     $\hat{B} \leftarrow$ CNN model($I$)
6: **while** $\hat{u}^N$ does not pass CRC && $t \leq T_{max}$ **do**
7:     $L, R \leftarrow$ initialize the BP decoder using (2)
8:     $i \leftarrow$ index of the $t$-highest value in $\hat{B}$ and mapped
       to the corresponding position of information bit
9:     $R_{0,i}^{(1)} \leftarrow \infty \times (2\hat{u}_i^N - 1)$
10:    $\hat{u}^N \leftarrow$ BP decoder($L, R$)
11:    $t \leftarrow t + 1$

that avoids updating the weights of part nodes, is also utilized to improve the prediction accuracy.

For the problem of bit-flipping prediction, the output layer has $K$ nodes, which represents the probability of each bit being flipped or not. The labeled data for training is a vector with $K$ values being 0 or 1 to indicate which bits could be flipped to result in successful decoding as shown in Fig. 3(c). Note that for some input cases, there could be more than 1 position to result in successful decoding. Consequently, this is a multi-label classification problem and the output must be rescaled into the range [0,1] with sigmoid function to indicate the probability as below:

$$f_{Sigmoid}(x) = \sigma(x) = (1 + e^{-x})^{-1}. \qquad (7)$$

Also, the loss function is cross entropy, is defined as:

$$\mathcal{L}(B, \hat{B}) = -\frac{1}{K} \sum_{i=1}^{K} B_i \log(\hat{B}_i) + (1 - B_i)\log(1 - \hat{B}_i), \qquad (8)$$

where $B_i$ and $\hat{B}_i$ denote the labeled data and predicted value for the $i$-th output, respectively.

Now, with the well-trained CNN model, the mechanism of proposed CNN-aided bit-flipping (CNN-BF) can commence as provided in Algorithm 1. The received signal will first go through its first round of BP decoding. After a pre-set number of iterations for BP, the CRC will be utilized to check whether the BP decoding is successful. If not, the mechanism of CNN-BF takes the transformed metadata from BP as input and outputs $K$ values to indicate the probability of whether each bit should be flipped. Then, the order of priority to attempt BF is based on the magnitude of probability as shown in Fig. 3(c). Starting from the highest probability, the bits will be attempted in order by likelihood until the CRC is passed or the maximum trials $T_{max}$ in bit flipping is reached.

TABLE I. SIMULATION PARAMETERS

| Encoding | Polar code (64,32) |
|---|---|
| Decoding Algorithm | RNN-BP [8] |
| Number of BP Iteration | 5 |
| Signal to Noise Ratio (SNR) | 0, 1, 2, 3 |
| CRC Generator Polynomial | $x^6 + x^5 + 1$ |
| Training Codeword/SNR | 38,400 |
| Testing Codeword/SNR | 153,600 |
| Validation Ratio | 0.2 |
| Mini-batch Size | 500 |
| Optimizer | Adam |
| Training and Testing Environment | DL library of Keras with NVIDIA RTX 8000 GPU |

TABLE II. COVERAGE RATE OF CRITICAL SET

| SNR (dB) | 0 | 1 | 2 | 3 |
|---|---|---|---|---|
| CS-BF [15] | 76.68% | 79.50% | 83.25% | 87.81% |

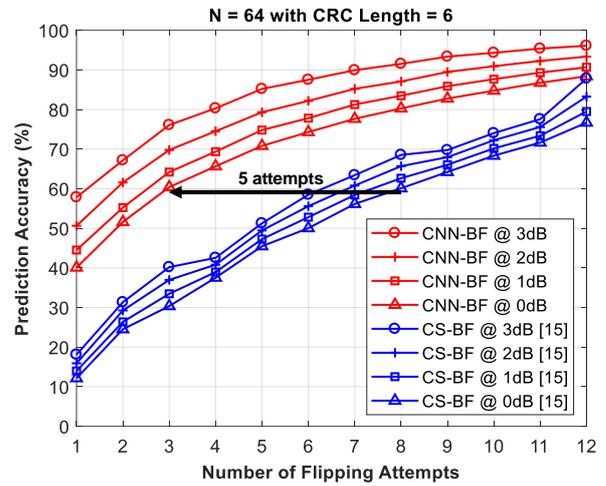

Fig. 4. Comparison of prediction accuracy between the proposed CNN-BF and CS-BF [15] under the different number of flipping attempts.

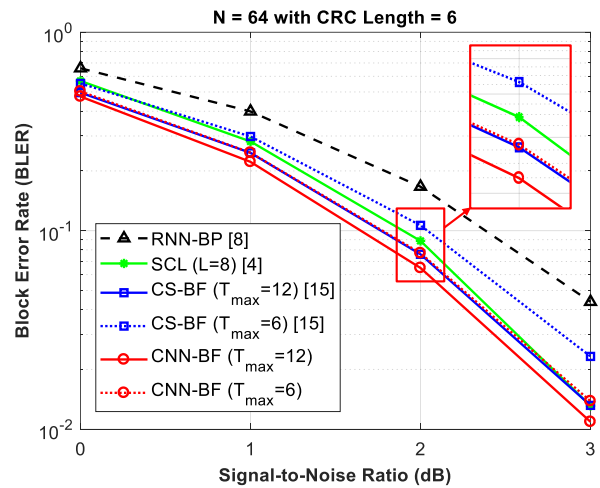

Fig. 5. Comparison of block error rate between the proposed CNN-BF, CS-BF [15], RNN-BP [8], and SCL [4] under different SNR.

## IV. SIMULATION RESULTS

In this work, we utilize the recurrent neural network-based belief propagation (RNN-BP) algorithm [8] to replace the conventional BP decoding algorithms. The RNN-BP can dramatically reduce the required number of BP iterations from 40 to 5, which decreases the additional decoding latency caused by each flipping attempt and makes the BF mechanism more practical. The simulation setup is summarized in Table I.

### A. Error Correction Capability of CNN-BF and CS-BF

As mentioned in Section III.A, though critical set can effectively reduce the search space for flipping attempt, it also excludes error bits outside the critical set, thus degrading the error correction capability of BF. The first experiment is to evaluate the gap between CS-BF and our proposed CNN-BF. Because the search space for CNN-BF is $K$, it can cover all error bits. On the other hand, the search space for CS-BF is limited to $|CS|$, which is 12 when $N = 64$. The evaluation of coverage rate for CS-BF is listed in Table II. From Table II, we can observe that though $|CS|$ is far less than $K$, it still can cover most of the error bits, which demonstrates the benefit of critical set for decreasing the search space. However, it also reveals the deficiency of CS-BF for degraded error correction capacity, especially for low signal to noise ratio (SNR) conditions.

### B. Comparison of Prediction Accuracy between CNN-BF and CS-BF

Next, we evaluate the performance of the proposed CNN-BF and CS-BF. The training and testing SNR is set from 0 dB to 3 dB as listed in Table I. First, we compare the BF selection accuracy between both methods. The number of flipping attempts is the number of tries until the decoding result passes CRC. The accuracy is determined by the number of cumulative successful decodings at the number of flipping attempts as a percentage of the total samples. The maximum number of flipping attempts $T_{max}$ is set to 12 which is as same as $|CS|$.

As seen in Fig. 4, both methods have better prediction accuracy as the SNR increases. However, CNN-BF predicts the correct bit-flipping condition at a significantly better accuracy, especially at earlier number of attempts, as well as having a higher ceiling for improvement over numerous bit-flipping attempts. These outstanding improvements are the result of two reasons. First, the well-trained CNN model has a more accurate BF selection. Second, CNN-BF can flip bits outside of the critical set which achieves better error correction capability over CS-BF. Both of the reasons contribute to the reduction of 5 flipping attempts for CNN-BF compared to CS-BF, which can effectively reduce the latency of the decoding process caused by bit-flipping.

### C. Comparison of Block Error Rate between CNN-BF and Prior Works

To further quantify the above results, we realize the contribution of successful decoding to the block error rate. The simulation results are shown in Fig. 5. In addition to the performance of CNN-BF and CS-BF, we also compare the performance of RNN-BP without BF mechanism to evaluate the improvement. The performance of SCL with list size $L = 8$ is also included [4]. Also, we compare the decoding performance when $T_{max} = 6$ and $T_{max} = 12$, respectively, where $T_{max}$ denotes the maximum trials in bit flipping. From Fig. 5, we can observe that SCL has better performance than RNN-BP due to the most likely paths are kept to avoid the mistakes happened in early stages. However, it suffers from high latency and low throughput due to its sequential processing nature. On the other hand, both CNN-BF and CS-BF, based on BP decoding algorithm, also achieve great improvement at the sacrifice of slightly longer decoding latency, which demonstrates that the BF mechanism can provide a compromise for adjustment between decoding performance and latency. Furthermore, CNN-BF can even outperform SCL when $T_{max}$ is set to 6, which is half of the 12 set for CS-BF due to its higher prediction accuracy and better error correction capability as shown in Fig. 4.

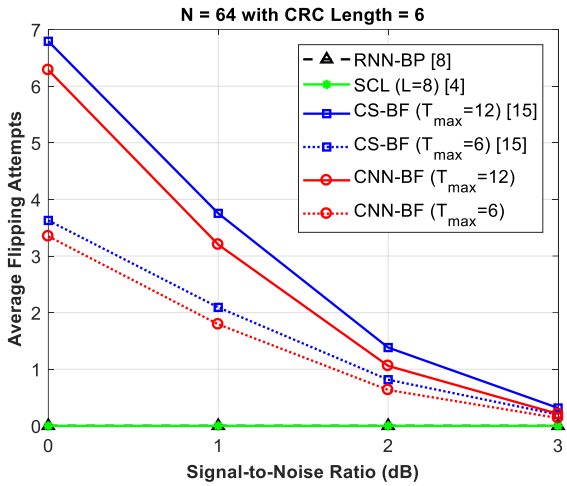

Fig. 6. Comparison of average flipping attempts between the proposed CNN-BF, CS-BF [15], RNN-BP [8], and SCL [4] under different SNR.

### D. Comparison of Average Flipping Attempts between CNN-BF and Prior Works

Finally, we examine the average flipping attempts between different approaches to evaluate the impact of additional decoding latency caused by bit-flipping. Besides, $T_{max} = 6$ and $T_{max} = 12$ are also compared. From Fig. 6, the average flipping attempts decreases rapidly as SNR increases. Especially, at SNR = 3dB and $T_{max} = 12$, the flipping attempts for CNN-BF and CS-BF are merely 0.21 and 0.32, respectively. It represents that the increase in decoding latency is small enough. However, it still contributes to significant improvement in decoding performance as shown in Fig. 5. Besides, under the same decoding performance when $T_{max}$ is 6 and 12 for CNN-BF and CS-BF, respectively, the average flipping attempts of CNN-BF is also half that of CS-BF in the entire SNR range. In summary, compared with CS-BF, the proposed CNN-BF not only achieves better decoding performance but also reduces the decoding latency due to the appropriate input data and well-designed network model.

## V. CONCLUSION

In this paper, we present a novel convolutional neural network-aided bit-flipping decoder. With carefully designed input data and domain-specific data pre-processing, our model can learn from BP metadata to correctly predict flipping position, with more accuracy than the prior critical set method. Therefore, it can avoid incorrect bit-flipping attempts with reduction in both decoding latency and error rate. Meanwhile, it also provides a flexible adjustment between decoding performance and latency, which fits into various requirements in 5G and future-generation communications.